\documentclass[prb,showpacs,amsmath,amssymb,twocolumn]{revtex4}
\usepackage{amssymb}
\usepackage{graphicx}
\usepackage{dcolumn}
\usepackage{bm}
\newcommand{\PrOsAs}{PrOs$_{4}$As$_{12}$}
\newcommand{\PrOsP}{PrOs$_{4}$P$_{12}$}

\newcommand{\LaFeP}{LaFe$_{4}$P$_{12}$}
\newcommand{\PrRuP}{PrRu$_{4}$P$_{12}$}

\newcommand{\PrFeP}{PrFe$_{4}$P$_{12}$}

\newcommand{\PrOsSb}{PrOs$_{4}$Sb$_{12}$}

\begin{document}

\title{Multiple ordered phases in the filled skutterudite compound  \PrOsAs{}}
\author{W.~M.~Yuhasz}
\author{N.~P.~Butch}
\author{T.~A.~Sayles}
\author{P.-C.~Ho}
\author{J.~R.~Jeffries}
\author{T.~Yanagisawa}
\author{N.~A.~Frederick}
\author{M.~B.~Maple}
\affiliation{Department of Physics and Institute for Pure and
Applied Physical Sciences, University of California San Diego, La
Jolla, CA 92093}

\author{Z.~Henkie}
\author{A.~Pietraszko}
\affiliation{Institute of Low Temperature and Structure Research,
Polish Academy of Sciences, 50-950 Wroc{\l}aw, Poland}

\author{S.~K.~McCall}
\author{M.~W.~McElfresh}
\author{M.~J.~Fluss}
\affiliation{Lawrence Livermore National Laboratory, P.O. Box 808,
Livermore, CA 94550}

\date{\today}

\begin{abstract}
Magnetization, specific heat, and electrical resistivity
measurements were made on single crystals of the filled skutterudite
compound \PrOsAs{}. Specific heat measurements indicate an
electronic specific heat coefficient $\gamma$ $\sim
50-200$~mJ/mol~K$^{2}$ at temperatures $10$~K~$\leq T \leq 18$~K,
and $\sim 1$~J/mol~K$^{2}$ for $T \leq 1.6$~K. Magnetization,
specific heat, and electrical resistivity measurements reveal the
presence of two, or possibly three, ordered phases at temperatures
below $\sim 2.3$~K and in fields below $\sim 3$~T. The low
temperature phase displays antiferromagnetic characteristics, while
the nature of the ordering in the other phase(s) has yet to be
determined.
\end{abstract}


\maketitle

The filled skutterudite compounds, with the formula $MT_{4}X_{12}$
($M$ = alkali metal, alkaline-earth, lanthanide, or actinide; $T$
= Fe, Ru, or Os; and $X$ = P, As, or Sb), display a wide variety
of strongly correlated electron phenomena.\cite{Maple03a,Sales03,
Aoki05} Of particular interest are the Pr-based filled
skutterudites. The physical properties of these compounds are
dominated by the ground state and low lying excited state of the
Pr$^{3+}$ ion in the crystalline electric field (CEF), and the
hybridization of the Pr 4f-states with the ligand-states of the
surrounding Sb ions that compose the atomic cage within which each
Pr$^{3+}$ ion resides. A variety of correlated electron phenomena
have been observed in the Pr-based filled skutterudites:
conventional (BCS) and unconventional superconductivity, magnetic
order, quadrupolar order, metal-insulator transitions, Kondo
phenomena, heavy fermion behavior, and non-Fermi liquid behavior.
A prime example of this diversity is seen in the compound
\PrFeP{}, which undergoes a transition to an ordered state below
$6.5$~K that was originally thought to be antiferromagnetic (AFM)
in nature,\cite{Torikachvili87} but was later identified with
antiferroquadrupolar (AFQ) order.\cite{Hao03}  The suppression of
this AFQ ordering in magnetic fields results in the formation of a
heavy Fermi liquid state above the quadrupolar quantum critical
point (QCP).~\cite{Matsuda00, Sugawara02, Aoki02} Below $1$~K, a
possible high field ordered phase between $8$~T and $12$~T has
also been observed in a limited angular range around the [$111$]
direction.\cite{Tayama04} One of the most interesting Pr-based
materials is the compound \PrOsSb{}, which exhibits unconventional
superconductivity, involving heavy fermion quasiparticles, below a
critical temperature $T_{c}$  =
$1.85$~K.~\cite{Maple01,Bauer02,Maple02} The superconducting state
breaks time reversal symmetry,\cite{Aoki03} appears to consist of
several distinct superconducting phases,\cite{Maple02, Izawa03,
Cichorek05} and may have point nodes in the energy
gap,\cite{Maple03b,Izawa03,Chia03} while between $4.5$~T and
$16$~T and below $\sim 1$~K, an ordered phase is observed that has
been identified with AFQ order.\cite{Maple02, Maple03b, Ho03,
Kohgi03} This suggests that the extraordinary normal and
superconducting properties of \PrOsSb{} may be associated with the
proximity to a field-induced quadrupolar QCP, in accord with
previous conjectures.\cite{Maple01, Bauer02, Maple02}

\begin{table*}
\caption{Atomic coordinates, displacement parameters, and occupancy
factors for PrOs$_4$As$_{12}$.  U$_{eq}$ is defined as one third of
the trace of the orthogonalized U$_{ij}$ tensor.}
\begin{tabular}{|c|ccc|c|c|}
\hline &&&&&Occupancy\\
Atom&x&y&z&U$_{eq}$ (\AA$^{2}\times 10^{3}$)&factor\\
\hline Os & 0.25 & 0.25 & 0.25 & 4(1) & 1.00(2)\\
As & 0 & 0.3485(1) & 0.1487(1) & 7(1) & 0.96(2)\\
Pr & 0 & 0 & 0 & 16(1)& 0.97(2)\\
\hline
\end{tabular}\label{xraydata}
\end{table*}

Unlike the Pr-based filled skutterudite phosphides and antimonides,
the arsenides have not been investigated in much detail. These
materials provide an opportunity to discover other examples of
strongly correlated electron behavior displayed by the Pr-based
filled skutterudites. In this report, we present an investigation of
the physical properties of \PrOsAs{} single crystals. These
measurements reveal that \PrOsAs{} is a Kondo lattice compound with
an enhanced electronic specific heat coefficient $\gamma$ of $\sim
50-200$~mJ/mol~K$^{2}$ at temperatures $10$~K~$\leq T \leq 18$~K,
and $\sim 1$~J/mol~K$^{2}$ for $T \leq 1.6$~K. Multiple ordered
phases are observed below $2.3$~K and $3.2$~T, at least one of which
appears to be AFM in nature.

\begin{figure}[tbp]
    \begin{center}
    \includegraphics[width=3.375 in]{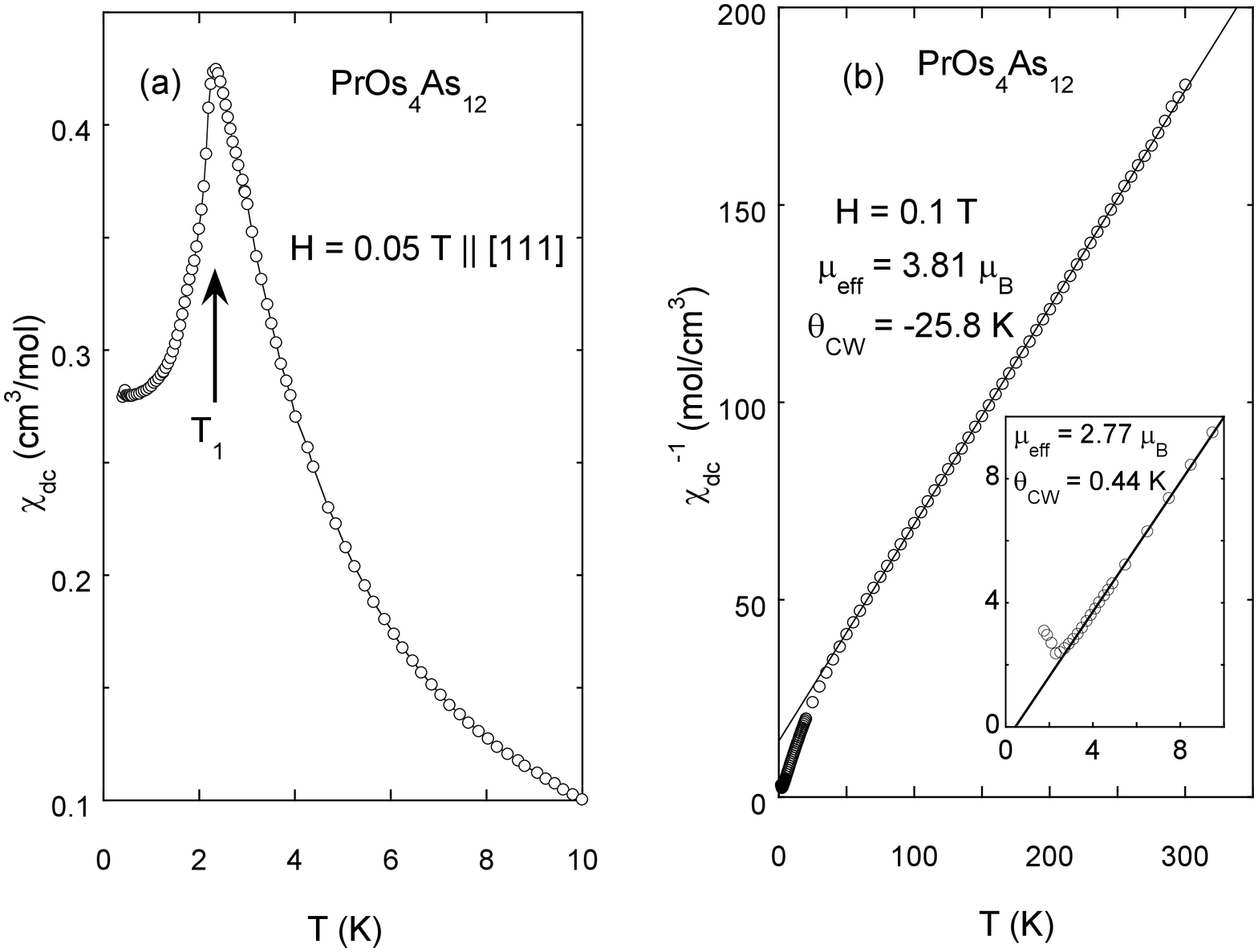}
    \end{center}
    \caption{(a) Magnetic susceptibility $\chi_\mathrm{dc}$ vs $T$, indicating a magnetic
    transition at $2.3$~K. (b) Inverse magnetic susceptibilility
    $\chi_\mathrm{dc}^{-1}$ vs $T$. The line represents a Curie-Weiss
    law, fit to data above $50$~K. The inset shows the low-$T$ behavior of $\chi_\mathrm{dc}^{-1}(T)$
    along with a different Curie-Weiss law fit (see text).}
    \label{X(T)}
\end{figure}

Single crystals of \PrOsAs{} were grown from elements with
purities $\geq$ 99.9\% by a molten metal flux method at high
temperatures and pressures, the details of which will be reported
elsewhere.\cite{Henkie06} After removing the majority of the flux
by distillation, \PrOsAs{} single crystals of an isometric form
with dimensions up to $\sim 0.7$~mm were collected and cleaned
further in acid to remove any Pr impurity phases from the
surfaces. X-ray powder diffraction measurements were performed
with a Rigaku D/MAX B X-ray machine on a powder prepared by
grinding several single crystals along with a high purity Si(8N)
standard. The crystal structure of \PrOsAs{} was determined by
X-ray diffraction (XRD) on a crystal with a regular octahedral
shape and dimensions of $0.12 \times{} 0.12 \times{} 0.12$~mm. A
total of $5158$ reflections ($583$ unique, R$_\mathrm{int} =
0.1249$) were recorded and the structure was resolved by the full
matrix least squares method using the SHELX-$97$ program with a
final discrepancy factor R$1 = 0.0462$ (for I~$> 2\sigma$(I), wR$2
= 0.1124$).\cite{Sheldrick85,Sheldrick87}

Magnetization $M$ measurements for temperatures $0.4$~K~$\leq T
\leq10$~K and magnetic fields $-5.5$~T~$\leq H \leq 5.5$~T were made
in a $^{3}$He Faraday magnetometer. A mosaic of $27$ crystals, with
a total mass of $27$~mg, was measured with the [111] axis of the
crystals nominally aligned parallel to $H$. Measurements of $M(T,H)$
were also performed using a Quantum Design Magnetic Properties
Measurement System (MPMS) in $H$ up to $5.5$~T for $1.7$~K~$\leq T
\leq 300$~K. Four-wire ac electrical resistivity $\rho(T)$
measurements were made for $2$~K~$\leq T \leq 300$~K in a Quantum
Design Physical Properties Measurement System (PPMS) using a
constant current of $1-10$~mA, and for $0.06$~K~$\leq T \leq 2.6$~K
in a $^{3}$He-$^{4}$He dilution refrigerator with a constant current
of $300~\mu$A. Measurements of $\rho(T)$ at various pressures $P$ up
to $\sim 23$~kbar were made down to $1$~K in a $^{4}$He cryostat
using a BeCu piston-cylinder clamp. Specific heat $C(T)$
measurements on \PrOsAs{} were performed for $0.6$~K~$\leq T \leq
35$~K at UCSD in a semi-adiabatic $^{3}$He calorimeter on a
collection of single crystals with a total mass of $49$~mg.
Additional $C(T)$ measurements for $0.4$~K~$\leq T \leq 20$~K were
performed at Lawrence Livermore National Laboratory (LLNL) in a
Quantum Design $^{3}$He PPMS.

Analysis of the X-ray powder diffraction pattern indicate single
phase \PrOsAs{} with no major impurity peaks.  A unit cell
parameter $a = 8.5319(11)$~\AA{} was determined, in excellent
agreement with an earlier measurement\cite{Braun80} of $a =
8.5311(3)$~\AA{}. Single crystal structural refinement shows that
the unit cell of \PrOsAs{} has the \LaFeP{}-type structure
(Im\={3} space group) with two formula units per unit cell, and
$a=8.520(1)$~\AA{}.  Other crystal structure parameters are
summarized in Table~\ref{xraydata}. The displacement parameter
U$_{eq}$ represents the average displacement of atoms vibrating
around lattice positions and equals mean-square displacements
along the cartesian axes. The displacement parameters determined
for \PrOsAs{} show behavior which is typical of the lanthanide
filled skutterudites.\cite{Sales97,Sales99}  Table~\ref{xraydata}
also indicates that the Pr site in \PrOsAs{} can be assumed to be
fully occupied since there is a $\sim 2\%$ uncertainty in the
occupancy factors.

\begin{figure}[tbp]
    \begin{center}
    \rotatebox{0}{\includegraphics[width=3.375in]{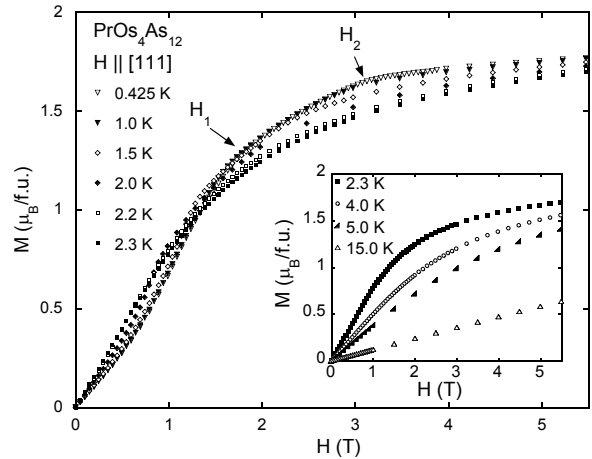}}
    \end{center}
    \caption{Magnetization $M$ vs $H$ isotherms at several values of $T$. Features in
    $M(H)$ at constant $T$, with $H_{1}$ and $H_{2}$ defined by a kink in $M(H)$, are illustrated for $T = 0.425$~K.
    The inset shows $M(H)$ at higher temperatures.}
    \label{MHcompare}
\end{figure}

The dc magnetic susceptibility $\chi_\mathrm{dc}(T)$, measured in $H
= 0.05$~T along the [111] direction, for $0.4$~K~$\leq T \leq 10$~K,
is shown in Fig.~\ref{X(T)}(a). The clear peak in the
$\chi_\mathrm{dc}(T)$ data suggests an AFM transition at $T =
2.3$~K. Figure~\ref{X(T)}(b) displays $\chi_\mathrm{dc}^{-1}$ vs $T$
in $H=0.1$~T between $2$~K and $300$~K. Above $50$~K, the data are
well-described by a Curie-Weiss law with an effective moment
$\mu_\mathrm{eff} = 3.81 \mu_\mathrm{B}$/f.u.\ and a Curie-Weiss
temperature $\Theta_\mathrm{CW} = -25.8$~K.  Below $20$~K,
$\chi_\mathrm{dc}^{-1}(T)$ can be described by a different
Curie-Weiss law, with $\mu_\mathrm{eff}$ = $2.77
\mu_\mathrm{B}$/f.u.\ and $\Theta_\mathrm{CW} = 0.44$~K
[Fig.~\ref{X(T)}(b) inset]; this behavior is consistent with
crystalline electric field (CEF) effects, as will be discussed
later.  The value of $\mu_\mathrm{eff}$ derived from the high
temperature (above $50$~K) Curie-Weiss fit is larger than the
theoretical Pr$^{3+}$ free-ion value of $\mu_\mathrm{eff} = 3.58
\mu_\mathrm{B}$/f.u.  A similar $\mu_\mathrm{eff} = 3.84
\mu_\mathrm{B}$/f.u. was also found for \PrRuP{}.\cite{Sekine97} In
contrast, \PrOsSb{} displays an effective moment of $2.97
\mu_\mathrm{B}$/f.u.,\cite{Bauer02} while both \PrOsP{} and \PrFeP{}
display effective moments close to the Pr$^{3+}$ free-ion
value.\cite{Sekine97,Torikachvili87} The larger-than-expected moment
may be due to an unaccounted-for constant Pauli paramagnetic
contribution. When a constant contribution $\chi_{0}$ of $\sim 5.5
\times 10^{-4}$~cm$^{3}$/mol is assumed, which is consistent with a
value of $5.16 \times 10^{-4}$~cm$^{3}$/mol found for
\LaFeP{},\cite{Meisner84} the effective moment estimated from a
Curie-Weiss law fit approaches the Pr$^{3+}$ free-ion value with
$\Theta_\mathrm{CW} = -12.5$~K.  The values of $\mu_\mathrm{eff}$
and $\Theta_\mathrm{CW}$ for the low temperature (below $20$~K)
Curie-Weiss law fit remain unchanged when the $\chi_\mathrm{dc}(T)$
data are corrected for this value of $\chi_{0}$.

Low-$T$ $M(H)$ data taken with $H$ parallel to [111] are shown in
Fig.~\ref{MHcompare}. An inflection point at low $H$ is evident,
along with two kinks in $M(H)$ (denoted $H_{1}$ and $H_{2}$) which
are suppressed toward lower field with increasing $T$. The inset of
Fig.~\ref{MHcompare} shows the evolution of $M(H)$ from higher
temperatures down to the ordering transition temperature. No
magnetic hysteresis was observed. Features corresponding to $H_{1}$
and $H_{2}$ are found in $M(T)$ in constant $H$.  With increasing
$H$, a peak in $M(T)$ [labeled as $T_\mathrm{1}$ in
Fig.~\ref{X(T)}(a)] shifts towards $T = 0$~K, while a kink in $M(T)$
at higher $T$ (denoted $T_\mathrm{2}$) becomes apparent, splits off
from $T_\mathrm{1}$, and eventually also moves towards $T = 0$~K.

\begin{figure}[tbp]
    \begin{center}
    \includegraphics[width=3.375in]{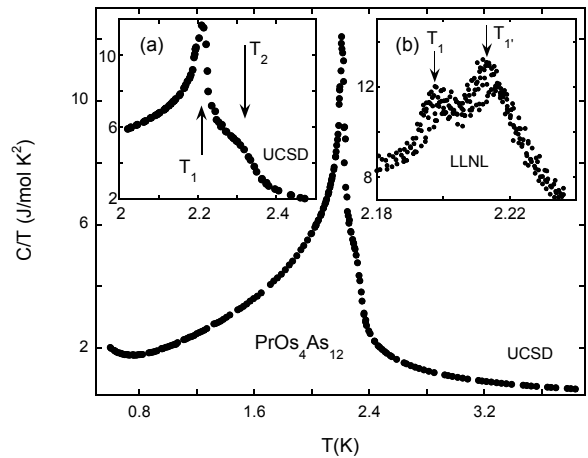}
    \end{center}
    \caption{Specific heat $C$ divided by $T$ vs $T$ (UCSD data).  Inset (a) shows the transitions
    at $T_\mathrm{1}$ and $T_\mathrm{2}$ in more detail (UCSD data).
    Inset (b) shows the apparent resolution of the transition at $T_\mathrm{1}$ into two transitions at $T_\mathrm{1}$
    and $T_\mathrm{1}$' (LLNL data).}
    \label{zerofieldCT}
\end{figure}

\begin{figure}[tbp]
    \begin{center}
    \includegraphics[width=3.375in]{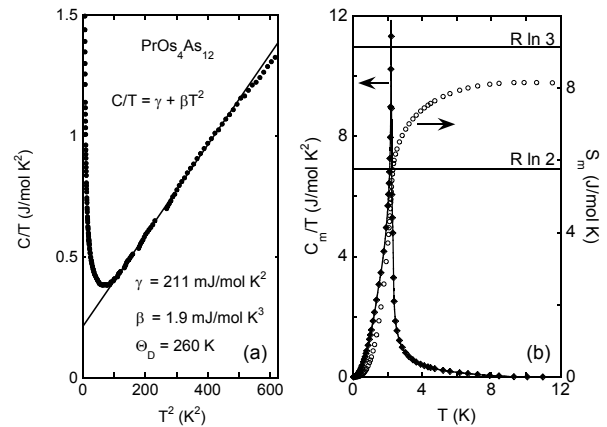}
    \end{center}
    \caption{(a) Zero field $C/T$ vs $T^{2}$, with a linear best fit yielding
    the enhanced $\gamma$ and value of $\Theta_D$ given in the figure.
    (b) Magnetic entropy $S_{m}$ determined from the specific heat after the
    zero field lattice and electronic contributions to the specific heat were subtracted.}
    \label{Smag}
\end{figure}

Specific heat $C(T)$ measurements on PrOs$_{4}$As$_{12}$ in zero
field (Fig.~\ref{zerofieldCT}) display two clear features: a slight
shoulder at $T_{2}\approx 2.3$~K, on the high temperature side of a
well defined peak at $T_{1}\approx 2.2$~K. Zero-field measurements
made at LLNL show similar behavior [Fig.~\ref{zerofieldCT} inset
(b)], except that in the LLNL experiment, the peak at $T_{1}$ was
further resolved into two distinct features separated by $\sim
0.02$~K.  This additional feature may indicate an additional
transition or it may be due to a variation in composition among the
measured crystals. At the lowest temperatures, there is a slight
upturn in $C(T)$ associated with a nuclear Schottky anomaly. The
low-$T$ specific heat can be described by four contributions,
$C(T)=C_\mathrm{n}(T)+C_\mathrm{e}(T)+C_{l}(T)+C_\mathrm{m}(T)$,
which are the nuclear Schottky, electronic, lattice, and magnetic
terms, respectively.  At higher temperatures, $C_\mathrm{m}(T)$ and
$C_\mathrm{n}(T) \approx 0$, and the specific heat is dominated by
the electronic and lattice terms. In Fig.~\ref{Smag}(a), a fit of
$C/T = \gamma + \beta T^{2}$ in the range $10$~K~$\leq T \leq 18$~K
reveals an enhanced $\gamma \approx 200$~mJ/mol K$^{2}$ and, from
$\beta$, a Debye temperature $\Theta_{D}\approx 260$~K. To analyze
the specific heat in the ordered state, the three terms of the
specific heat after subtraction of the lattice contribution (using
$\Theta_{D}\approx 260$~K), $\Delta C(T) = C_\mathrm{n} +
C_\mathrm{e} + C_\mathrm{m}$, were assumed to have the following
forms: $C_\mathrm{n}(T) = A/T^{2}$, $C_\mathrm{e}(T) = \gamma{}T$,
and $C_\mathrm{m}(T) = BT^{n}$. Using these equations, fits of
$\Delta{}CT^{2} = A + \gamma T^{3} + BT^{n+2}$ were performed for $T
< 0.6 T_{1}$. The best fit results in a nuclear Schottky coefficient
$A = 128$~mJ~K/mol and an exponent in the power-law term of $n = 3.2
\pm 0.1$, close to the expected value of $n = 3$ for AFM spin waves.
The electronic specific heat coefficient $\gamma$ is strongly
enhanced, with a value near $1$~J/mol~K$^{2}$. A temperature
dependent $\gamma$ is not uncommon and has been observed in numerous
compounds that exhibit heavy fermion behavior.\cite{Maple01b} The
value of $260$~K for $\Theta_{D}$ is consistent with other filled
skutterudite arsenides with typical values ranging from $230$~K
(LaRu$_{4}$As$_{12}$) to $340$~K
(PrRu$_{4}$As$_{12}$).\cite{Shirotani97,Yuhasz06} The magnetic
entropy $S_\mathrm{m}(T)$ was calculated from the specific heat by
integrating $C_\mathrm{m}(T)/T$ with temperature
[Fig.~\ref{Smag}(b)], resulting in an entropy release of $90\%$
$R\ln 3$ that levels off by $8$~K, a value consistent with a triply
degenerate ground state.  The reduction of the entropy from a full
$R\ln 3$ is presumably due to hybridization of the localized 4f and
itinerant electron states, as is observed for \PrOsSb{}, where
entropy is apparently transferred from the localized 4f states to
the conduction electron states.

\begin{figure}[tbp]
    \begin{center}
    \includegraphics[width=3.375in]{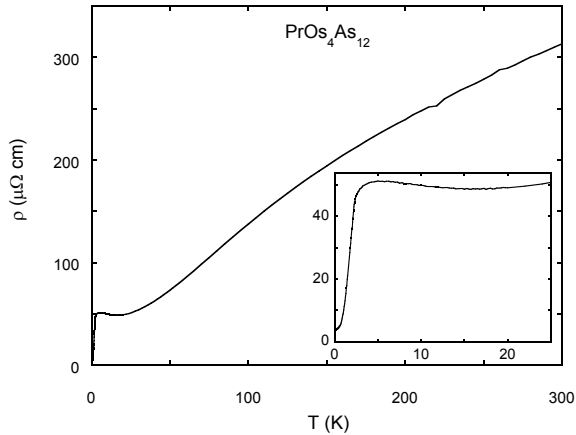}
    \end{center}
    \caption{Electrical resistivity $\rho$ vs $T$ from
    $2$~K to $300$~K. The inset shows $\rho(T)$ from $0.06$~K to $25$~K.}
    \label{HiLoTRes}
\end{figure}

The electrical resistivity $\rho(T)$ was measured in the range
$0.06$~K~$\leq T \leq 300$~K (Fig.~\ref{HiLoTRes}). Kondo lattice
behavior is evident, as $\rho(T)$ decreases with decreasing $T$ down
to $T_\mathrm{min} \approx 16.5$~K and then increases until $\sim
5.5$~K, where the resistivity drops due to the onset of the ordered
state. Measurements in high magnetic fields can be analyzed in terms
of a single ion Kondo model for an impurity effective spin of $1$,
yielding an estimate of $T_{K}$ on the order of $1$~K.\cite{Maple06}
A small $T_{K}$ with respect to the ordering temperature($\sim
2.3$~K) prohibits the observation of $T^{2}$ dependence of
$\rho(T)$. Measurements of $\rho(T)$ under $P$ up to $\sim 23$~kbar
between $1$~K and $300$~K revealed no significant $P$ dependence.

As mentioned earlier, features in the $\chi_\mathrm{dc}(T)$ data
reflect CEF effects. Although the true crystal symmetry of the
filled skutterudites is tetrahedral $T_{h}$, it is a close
approximation in zero field to consider cubic $O_{h}$ symmetry. In a
CEF with $O_{h}$ symmetry, the ninefold degenerate Pr$^{3+}$ $J=4$
multiplet is split into a $\Gamma_{1}$ singlet, a $\Gamma_{3}$
doublet, and $\Gamma_{4}$ and $\Gamma_{5}$ triplets. The low-$T$
calculated effective moment of $2.77 \mu_\mathrm{B}$/f.u.\ is
indicative of a $\Gamma_{5}$ ground state, which has a theoretical
effective moment of $2.83 \mu_\mathrm{B}$/f.u. In addition, $M =
1.68 \mu_\mathrm{B}$/f.u.\ at $2$~K and $5.5$~T, a value much less
than the full $3.2 \mu_\mathrm{B}$/f.u.\ saturated moment expected
for a Pr$^{3+}$ free ion, but more consistent with the $2.0
\mu_\mathrm{B}$/f.u.\ expected for a $\Gamma_{5}$ ground state.
Reasonable CEF fits to the $\chi_\mathrm{dc}(T)-\chi_{0}$ data (not
shown) can be made using a CEF splitting as low as $\sim 15$~K
between a $\Gamma_{5}$ ground state and a $\Gamma_{1}$ first excited
state. If there were a large Schottky contribution due to this CEF
splitting in the temperature range of the high temperature fit to
$C(T)$, the values of $\gamma$ and $\Theta_{D}$ determined from the
fit would be altered. Based on the CEF fits to the
$\chi_\mathrm{dc}(T)-\chi_{0}$ data, the worst case scenario (which
would lead to the largest possible change in $\gamma$) occurs for a
splitting of $23$~K. The resulting Schottky contribution to the
specific heat would result in a reduced $\gamma$ ($50$~mJ/mol
K$^{2}$) and a slightly lower $\Theta_{D}$ ($246$~K). Since the CEF
contribution to C(T) is not significant below $1.6$~K for splittings
above $15$~K, the large value of $\gamma$ ($\sim 1$~J/mol K$^{2}$)
determined from the low temperature fit to $C(T)$ is not affected by
potential CEF effects. The various acceptable CEF splitting schemes
do not significantly alter the magnetic entropy release of $90\%$
$R\ln 3$ by $8$~K, which remains in support of a triplet ground
state.

\begin{figure}[tbp]
    \begin{center}
    \rotatebox{0}{\includegraphics[width=3.375in]{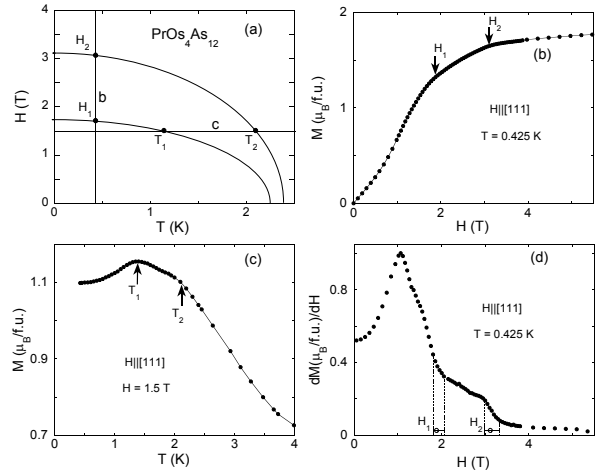}}
    \end{center}
    \caption{(a) Schematic representation of the \PrOsAs{} magnetic field - temperature
    $(H-T)$ phase diagram.  The vertical and horizontal lines represent isothermal $M(H)$ and
    constant field $M(T)$ measurements as depicted in (b) and (c). (b) Magnetization $M$ vs $H$
    at $0.425$~K, with features at $H_{1}$ and $H_{2}$ used to define the phase diagram.
    (c) Magnetization $M$ vs $T$ at $1.5$~T, along with two clear features at $T_{1}$ and
    $T_{2}$, used to contrast the phase diagram. (d)  The derivative $dM/dH$ as a function of $H$
    indicates two features used to define $H_{1}$ and $H_{2}$ in Fig.~\ref{MagFeatures}(b).
    The dotted lines represent estimated transition widths for $H_{1}$ and $H_{2}$.}
    \label{MagFeatures}
\end{figure}

\begin{figure}[tbp]
    \begin{center}
    \rotatebox{0}{\includegraphics[width=3.375in]{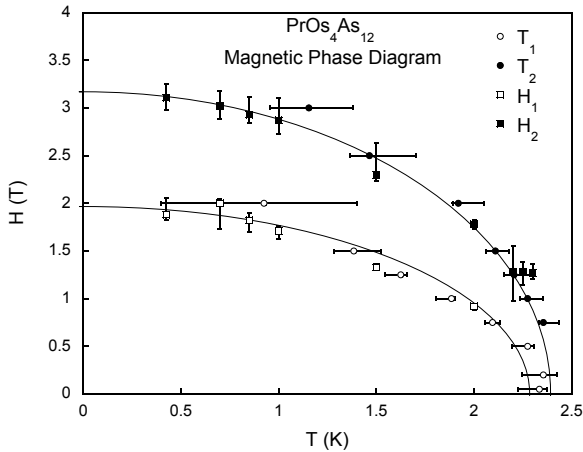}}
    \end{center}
    \caption{Magnetic field-temperature ($H-T$) phase diagram for \PrOsAs{}, as defined by features
    in the magnetization (Fig.~\ref{MHcompare})(circles and squares). The lines are
    guides to the eye.}
    \label{MagPhaseDiagram}
\end{figure}

A schematic representation of the $(H-T)$ phase diagram for
\PrOsAs{} (Fig.~\ref{MagPhaseDiagram}), is shown in
Fig.~\ref{MagFeatures}(a). The vertical line labeled ``b" in the
schematic phase diagram represents an $M(H)$ measurement at $T =
0.425$~K shown in Fig.~\ref{MagFeatures}(b). The points $H_{1}$
and $H_{2}$ along line b are features in the $M(H)$ data, which
are defined using the derivative $dM/dH$ shown in
Fig.~\ref{MagFeatures}(d) as a function of $H$. The horizontal
line labeled ``c" represents an $M(T)$ measurement at $H = 1.5$~T
(Fig.~\ref{MagFeatures}(c)). The features at $T_{1}$ and $T_{2}$
along c are defined from the $M(T)$ data shown in
Fig.~\ref{MagFeatures}(c). Similar features in the other $M(H,T)$
measurements were used to generate the full $H-T$ phase diagram
for \PrOsAs{} shown in Fig.~\ref{MagPhaseDiagram}. The error bars
are rough estimates of the transition widths, based upon the
widths of features in the derivatives of $M(H)$
(Fig.~\ref{MagFeatures}(d)) and $M(T)$ (not shown). It may be
inferred from the features in the UCSD $C(T)$ data that there are
two phase transitions. Thus, \PrOsAs{} appears to have an AFM
ground state, a second ordered state at intermediate $T$ and $H$,
and is paramagnetic at high $T$ and $H$.

The AFM nature of the ground state is supported by several
observations. The peak in $\chi_\mathrm{dc}(T)$ is a traditional
indication of the onset of AFM, which is also suggested by the
lack of hysteresis in $M(H)$, and is compatible with a CEF-split
$\Gamma_{5}$ magnetic triplet ground state. Furthermore, the
suppression of the ground state order with $H$ is consistent with
AFM order. The nature of the second ordered phase is difficult to
establish from the results of this study, but in light of the
low-$T$ behavior of other Pr-based filled skutterudites such as
\PrFeP{},\cite{Hao03} magnetic or quadrupolar ordering seem likely
possibilities. Recent neutron diffraction experiments on \PrOsAs{}
confirm that the low field ordered phase is
antiferromagnetic.\cite{Maple06}

To summarize, magnetization and specific heat measurements on
\PrOsAs{} indicate the existence of at least two distinct phase
transitions at temperatures below $2.3$~K and in fields below
$3$~T. The magnetic behavior of the low temperature phase is
consistent with AFM order. The electrical resistivity displays a
sharp decrease due to the onset of ordering and has features
consistent with Kondo lattice behavior. In the paramagnetic state,
specific heat measurements indicate an electronic specific heat
coefficient with $\gamma$ $\approx 50-200$~mJ/mol~K$^{2}$.  In the
antiferromagnetic state, analysis of $C(T)$ data yields a very
large $\gamma \sim 1$~J/mol~K$^{2}$ at low $T$ ($T \leq 1.6$~K).

Research at UCSD was supported by the U. S. Department of Energy
under Grant No.~DE-FG02-04ER46105, the U.S. National Science
Foundation under Grant No.~DMR 0335173, and the National Nuclear
Security Administration under the Stewardship Science Academic
Alliances program through DOE Research Grant No.
DE-FG52-03NA00068.

\end{document}